\documentclass[a4paper]{jpconf}
\usepackage{graphicx}
\usepackage{amsthm,amscd,amssymb}
\usepackage{wrapfig}
\usepackage{epic,eepic,pstricks}
\usepackage{mathrsfs}
\usepackage{rotating}
\usepackage{placeins}
\usepackage{amsmath}
\usepackage{amsfonts}
\usepackage{graphicx}
\usepackage{psfrag}
\usepackage[font=small,skip=-10pt]{caption}
\usepackage[breaklinks=true]{hyperref}  % backref linktocpage pagebackref
\usepackage{siunitx} % Permite poner unidades en el texto
\usepackage{pfnote} %Reset footnotes every page

\newtheorem{theorem}{Theorem}
\newtheorem{corollary}{Corollary}
\newtheorem{proposition}{Proposition}

\def\vs{\vspace{5mm}}

\def\M{{\mathcal M}}
\def\N{{\mathcal N}}
\def\C{{\mathcal C}}
\def\Khat{\widehat{K}}
\def\phat{\widehat{p}}
\def\that{T}

\newcommand{\bm}[1]{\mbox{\boldmath $#1$}}

\makeatletter
\let\c@proposition\c@theorem
\let\c@corollary\c@theorem
\let\c@lemma\c@theorem
\let\c@definition\c@theorem
\let\c@example\c@theorem
\makeatother

\usepackage{microtype} % Slightly tweak font spacing for aesthetics

\begin{document}
\title{Regularization of geodesics in static spherically
symmetric Kerr-Schild spacetimes}

\author{Pablo Galindo}

\address{Dept. de Geometr\'{\i}a y Topolog\'{\i}a, Universidad de Granada, Campus de Fuentenueva s/n, 18071 Granada, Spain.}

\ead{pablogsal@correo.ugr.es}

\author{Marc Mars}

\address{Inst. de F\'{\i}sica Fundamental y Matem\'aticas (IUFFyM), Universidad de Salamanca, Plaza de la Merced s/n 37008 Salamanca, Spain.}

\ead{Marc@usal.es}

\begin{abstract}
We describe a method to analyze causal geodesics in static and spherically symmetric
spacetimes of Kerr-Schild form which, in particular, 
allows for a detailed study of the geodesics in the vicinity of the central
singularity by means of a regularization procedure based
on a generalization of the McGehee regularization for 
the motion of Newtonian point particles moving in a power-law potential. The
McGehee regularization was used by
Belbruno and Pretorius \cite{belbruno2011dynamical}
to perform a dynamical system regularization of
the central singularity of the motion of massless test
particles in the Schwarzschild spacetime. Our generalization allows us
to consider causal (timelike or null) geodesics 
in any static and spherically symmetric spacetime of Kerr-Schild form.
As an example, we apply these results to causal geodesics in the Schwarzschild and Reissner-Nordstr\"om spacetimes.
\end{abstract}

\section{Introduction} 

The motion of freely falling particles in static spherically symmetric
spacetimes constitutes a basic topic in General Relativity, which has been studied
from several perspectives. In this paper we use dynamical system methods to analyze causal geodesics in static spherically symmetric spacetimes admitting a
Kerr-Schild structure (which, in this context, is a very mild restriction and 
includes virtually all cases of interest). We are thus able to analyze the behaviour of geodesics across Killing horizons and reach the singularity at $r=0$, when
one is present.  We are particularly interested in studying the geodesics in the vicinity of the central singularity. To that aim we generalize 
the McGehee regularization \cite{mcgehee1981double} of power-law Newtonian
potentials, already used by Belbruno and Pretorius 
\cite{belbruno2011dynamical}  to study null geodesics in the
Schwarzschild spacetime, to a general Newtonian potential. We also show that
the dynamics described by the Hamiltonian of causal  geodesics in a static,
spherically symmetric spacetimes of Kerr-Schild form is equivalent 
to the dynamics of a Newtonian point particle under the action of
a suitable central potential. The regularization
at the center introduces a collision manifold which is invariant
under the flow and which allows for a detailed description of the approach
of geodesics to the singularity. The condition
that the geodesics be future causal translates into the existence 
of excluded regions of the phase space portrait. The topological modification
of the phase space coming from the excluded regions is intimately linked to
the Penrose structure of the maximal extension of a Kerr-Schild patch of the
spacetime under consideration, thus allowing us to describe the global behaviour
of geodesics in the extended spacetime by means of a single two-dimensional
phase space. We illustrate this by studying causal geodesics
in the Schwarzchild and Reisner-Norsdtr\"om spacetimes. Further details
on the topics addressed in this paper can be found in 
\cite{galindo2014mcgehee}.

\section{Geodesic equations for a general stationary Kerr-Schild metric}

Throughout this paper, we will consider spacetimes
$\{ \M = \mathbb{R} \times (\mathbb{R}^3 \setminus \C),g \}$ where
$\C \subset \mathbb{R}^3$ is a closed subset such that
$\M$ is connected and $g$ is a Lorentzian metric of Kerr-Schild form 
\cite{kerr1965new}.
More specifically, let  $\{ x^{\alpha} \} = \{ \that, x^i \}$ ($\alpha,\beta,
\cdots = 0,1,2,4$ and  $i,j, \cdots =1,2,3$)  be Cartesian coordinates on 
$\mathbb{R} \times \mathbb{R}^3$ and endow
$\M$ with the Minkowski metric
$\eta = - d \that^2 + \delta_{ij} dx^i dx^j $. Let $\bm{K}$ be a smooth
one-form on $\M$ which is null with respect to the metric $\eta$ and
$h: \M \longrightarrow \mathbb{R}$ a smooth function.
The metric $g$ being of Kerr-Schild form means that it takes the form
%$g  = \eta  +h \bm{K} \otimes \bm{K}$.
\begin{align*}
g  = \eta  +h \bm{K} \otimes \bm{K}.
\end{align*} 
In addition we will assume $\C$ to be rotationally symmetric 
and $g$ to be static and spherically symmetric
\begin{equation*}
 g  = -dT^2 + d\vec{x} \cdot d \vec{x} + h(r) (dr - \sigma dT) \otimes (dr - 
\sigma dT) ,
\end{equation*}
where  $r=\sqrt{\vec{x}\cdot \vec{x}}$ and $\sigma=\pm 1$. The integrable Killing vector is $\partial_T$, which we note is timelike, null or spacelike depending
on whether $h<1$, $h=1$ or $h >1$. We choose the time orientation so that $\sigma K_\alpha:=(dr - \sigma dT)$ is future directed. The function $h$ takes the form $h(r)=\frac{2M}{r}$ in the Schwarzschild spacetime and   $h(r)=\frac{2M r+Q^2}{r^2}$ in the Reissner-Nordstr\"om spacetime. In any spacetime $(\M,g)$, affinely parametrized geodesics
are the solutions of the Hamilton equations of the Hamiltonian
\begin{equation*}
H=\frac{1}{2}(g^{-1})^{\alpha \beta} p_\alpha p_\beta
\end{equation*}
defined on the cotangent bundle of $\M$. In our case, and by
the use of the Killing conserved quantities, the following Hamiltonian describing the spatial part of the geodesic flow arises naturally
\begin{equation}
\label{KerrSchildHamilton}
H^{\prime} := H + \frac{1}{2} E^2 = 
\frac{1}{2} \left( \vec{p}^{\,2}- h \left( \vec{K} \cdot \vec{p} - E 
\Khat \right)^2 \right),
\end{equation} 
where we have written $K = \{\hat{K},\vec{K}\}$, $\bm{p} = \{ \phat,\vec{p} \, \}$,  $E := - \bm{p} (dT)$ and dot means scalar product with $\delta_{ij}$. 
$H'$ is now defined on the cotangent bundle of $\mathbb{R}^{3} \setminus \C$.
In addition to the Hamilton equation of $H'$, the geodesics also
satisfy the following ODE describing the temporal part of the geodesic flow
\begin{equation}
\left (  1 -  h \vec{K}^{\, 2}  \right ) \frac{d \that}{ds}
+ h \Khat \vec{K} \cdot \frac{d \vec{x}}{ds} = E,
\label{eq:conservedenergy}
\end{equation}
and which is simply the coordinate form of $g(u, \partial_T) = -E$, where $u$
is the geodesic velocity vector $u(s):= (\dot{T}(s),\dot{\vec{x}}(s))$. 
The Hamiltonian $H^{\prime}$ is still complicated. The following
theorem \cite{galindo2014mcgehee} allows us to replace it
by a simpler one.
\begin{theorem} \label{HamiltonT}
Let $x'(s)$ and $\tilde{x}(s)$ be trajectories of the respective Hamiltonians
\begin{align*}
H' &= \frac{1}{2} \vec{p}\, {}^2- \frac{h(|\vec{x}|)}{2}  \left( \frac{\vec{x}
\cdot \vec{p}}{|\vec{x}|}  - \sigma E  \right)^2  ,  & E \in \mathbb{R} 
\nonumber \\
\tilde{H}& =\frac{1}{2} \vec{p}\, {}^2-\frac{h(|\vec{x}|)}{2} 
\left(\frac{L^2}{|\vec{x}|^2}+ \mu \right), & L, \mu \in \mathbb{R}.
\end{align*}
Then $x'(s)=\tilde{x}(s)$ if and only if their respective initial values
satisfy
\begin{align*} 
&\tilde{x}_0= x'_0, \quad \quad
\hspace{-1pt}  \tilde{p}_0 =  p'_0 - h(|x'_0|) \left ( \frac{
x'_0 \cdot p'_0}{|x'_0|} - \sigma E \right ) 
\frac{x'_0}{|x'_0|},\\
&| x'_0 \times p'_0 |  =  |L|, \quad \quad
H'(x'_0, p'_0)  = \frac{1}{2} \left ( E^2 - \mu \right 
).
\end{align*}
Moreover, in that case, $\tilde{H}(\tilde{x}_0,\tilde{p}_0) = \frac{1}{2} 
\left ( E^2 - \mu \right )$.
\end{theorem}
Note that the value of $\mu$ in the theorem is arbitrary. However, when making
contact to the spacetime geodesics, the relevant values are
$\mu=1$ for timelike geodesics, $\mu=0$ for null geodesics and $\mu=-1$ for spacelike geodesics, cf. (\ref{KerrSchildHamilton}). It is interesting that the  Hamiltonian $\tilde{H}$ is independent of $\sigma$, so that it can describe the geodesics in $(\M,g)$ both for the case when $\bm{K}$ is future directed (plus sign) or past directed (negative sign). Moreover, the Hamiltonian
$\tilde{H}$ is a standard Hamiltonian in Newtonian mechanics for a point particle 
in a central potential. This is a substantial simplification over the original problem of solving the geodesic equations in a stationary and spherically symmetric spacetime of Kerr-Schild form, because we can exploit all the information known for trajectories of point
particles in Newtonian mechanics under the influence of a radial potential. Hence the equation for the trajectories are
\begin{align*}
\ddot{\vec{x}} &= - \frac{\partial V(|\vec{x}|)}{\partial \vec{x}}= 
\frac{\partial}{\partial \vec{x}} \left[ \frac{h(|\vec{x}|)}{2} \left(\frac{L^2}{|\vec{x}|^2} + 
 \mu \right) \right], \\ %\label{spatialgeodesic}\\
\vec{L}  & = \vec{x} \times \dot{\vec{x}},
%\label{angularmomentum} 
\end{align*}
which need to be supplemented with the energy conservation stated at the end
of Theorem \ref{HamiltonT},
\begin{equation*}%\label{conservedgeodesic}
\frac{1}{2} |\dot{\vec{x}}|^2 - \frac{h(|\vec{x}|)}{2} \left( \frac{L^2}{|\vec{x}|^2} + \mu \right) = 
\frac{1}{2} \left( E^2-\mu \right)  := \epsilon
%:=\frac{\dot{r}^2}{2}  + V_\text{eff}(r).
\end{equation*}
and with (\ref{eq:conservedenergy}), which becomes
\begin{equation*}
E = (1 - h) \dot{T} + h \sigma \frac{\vec{x} \cdot \dot{\vec{x}}}{|\vec{x}|}.
%\label{Ener}
\end{equation*}

Since we are interested in future directed causal geodesics we need to find the restrictions on the initial data which guarantee this. This is given in the
following proposition, cf.  \cite{galindo2014mcgehee}  
\begin{proposition}\label{excluded}
A geodesic starting at a point $(t_0,\vec{x}_0 \neq 0)$ is future causal 
iff $\dot{\vec{x}}_0$ satisfies
 \begin{align*}
& \mbox{if } h_0>1, \quad
\begin{aligned}
\sigma \dot{r}_0 \in [a_0, \infty)
\end{aligned}
\\
&\mbox{if } h_0<1,  \quad
 \begin{aligned}
&\dot{r}_0 = (-\infty,\infty) \\
\end{aligned}
\\
& \mbox{if } h_0=1,  \quad  
\begin{aligned}
& \sigma \dot{r}_0 \in [0, \infty)  \quad \quad \mbox{with} \quad
\dot{r}_0 =0 \Longrightarrow
\mu = L =0 \\
\end{aligned}
\end{align*}
where $r_0 := |\vec{x}_0|$, $\dot{r}_0 := \frac{\vec{x}_0 \cdot
\dot{\vec{x}}_0}{|\vec{x}_0|}$,
$h_0 := h(r_0)$ and 
$a_0 :=  \sqrt{\left ( |1- h_0| \right ) \left ( \frac{L^2}{r_0^2} 
+ \mu \right )} \geq 0 $.
\end{proposition}
These restrictions imply the existence of excluded regions in the phase
space:
\begin{corollary}\label{epsilonrange}
The excluded regions in the phase space correspond to
$h_0 \geq 1$ and $\epsilon < - \frac{\mu}{2}$,
independently of the sign of $\sigma$  and of the function
$h(|\vec{x}|)$ in the 
Kerr-Schild metric.
%variation ranges for $\epsilon$ are
%\begin{equation}
%\begin{cases}
%\epsilon \in [-\frac{\mu}{2},\infty) &\mbox{if } h_0 \geq 1\\
%\epsilon \in [\frac{a_0^2-\mu}{2},\infty) &\mbox{if } h_0<1
%\end{cases}
%\end{equation}
\end{corollary}

Having obtained the geodesic equations for all values of $\sigma$ we need a method that allows us to regularize the singularity located at $r=0$. The method that we develop in the following theorem, see \cite{galindo2014mcgehee},
is called ``McGehee regularization'' because it provides
a generalization of the original approach by McGehee in \cite{mcgehee1981double}. This procedure will allow us to obtain information of the geodesics
at the vicinity of the singularity
\begin{theorem}\label{Regularizationtheorem}
Let $\N$ be an open annulus in $\mathbb{C}$
  and $V : \N \rightarrow \mathbb{R}$ be a radially symmetric 
function $V(x) = V(|x|)$. Assume that  $V(|x|)$ is $C^1$ as 
a function of $|x|$ and define $\nabla = \partial_{x^1} + i \partial_{x^2}$
where $x = x^1+ i x^2$, $x^1,x^2 \in \mathbb{R}$. Then the
dynamical system
\begin{align}
\dot{x} &  = y, \nonumber \\
\dot{y} &  = -\nabla V(|x|):= \Lambda(|x|) x, \label{RadialDS}
\end{align}
on $\N \times \mathbb{C}$ 
is equivalent to the
system
\begin{align}
  r'&=r u \label{Generaldecoupled1}\\
u'&=r^{-2 (\beta +1)} \left(L^2+r^4 \Lambda (r)\right)-\beta  
u^2 \label{Generaldecoupled2} \\
 \theta'&=L r^{-(\beta +1)} \label{Generaldecoupled3}
\end{align}
where $\beta$ is
an arbitrary constant. This system also admits the following two constants of motion
\begin{align*}
L & = v r^{\beta +1} \\ %\label{constantL} \\
\epsilon & = \frac{1}{2} r^{2\beta} \left (u^2 + v^2 \right ) + V(r).
%\label{constantep}
\end{align*}
The coordinates $\{r,\theta,u,v\}$
take values in $r \in (a,b) \subset \mathbb{R^{+}}$,
$\theta \in \mathbb{S}^1$ and $u,v \in \mathbb{R}$. The coordinate change
is defined
by
 \begin{equation*}
\left . \begin{array}{ll}
x&  = \,  r e^{i \theta} \\ 
y&= \, r^{\beta} (u+ i v)e^{i \theta} 
\end{array} \right \} \quad \quad \quad \quad 
d\tau = r^{1-\beta} ds, 
\end{equation*}
where $\tau$ is the flow parameter in (\ref{RadialDS}) and
$s$ is the flow parameter in
(\ref{Generaldecoupled1})-(\ref{Generaldecoupled3}).
\end{theorem}
The optimal choice of $\beta$
for a detailed study of the dynamical
system (\ref{Generaldecoupled1})-(\ref{Generaldecoupled3}) at $r=0$ is
$\beta = \mbox{min} \{  -1 , - \frac{\gamma}{2} \}$
\cite{galindo2014mcgehee}  
with $\gamma$ selected in such a way that $|x|^{\gamma} V(|x|)$ admits a
$C^1$ extension to $|x|=0$ and $\lim_{|x| \rightarrow 0} |x|^{\gamma} V(|x|) \neq 0$.
Indeed, a larger value of $\beta$
is not capable of regularizing the system at $r=0$. On the
other hand, a smaller value of $\beta$
overkills the singularity. This has the effect that the invariant submanifold $\{r=0\}$
(which is called {\it the collision manifold}) has $u=0$ as a
single 
fixed point, which is moreover always non-hyperbolic. Thus, all details of the phase space
structure of the dynamical system at $\{r=0\}$ are lost by such a
choice of $\beta$. 
We will see below an example of this behavior when considering the Schwarzschild limit
of the dynamical system describing causal geodesics in the Reissner-Nordstr\"om spacetime.
\FloatBarrier

\section{The Schwarzschild dynamical system}
\begin{figure}
\centering
\includegraphics[scale=0.45]{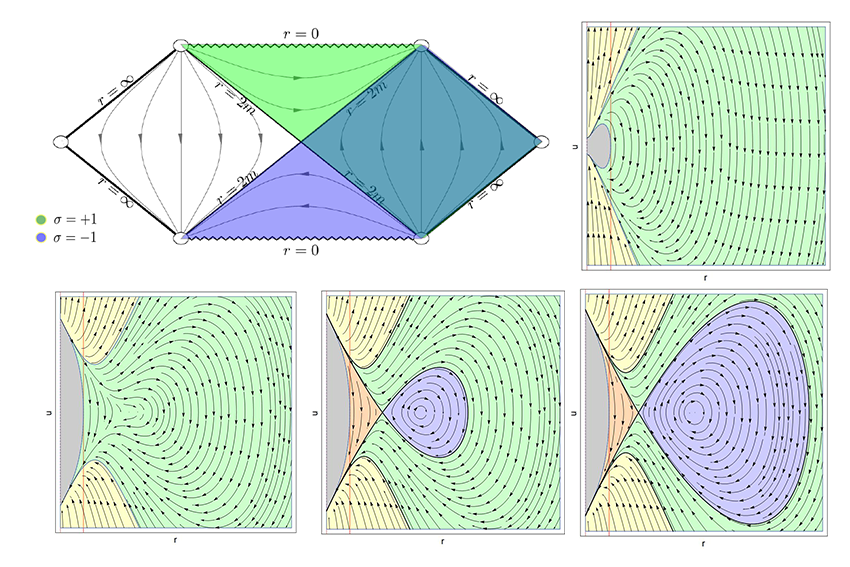}
\vspace*{1em}
\caption{Phase space for timelike particles with increasing value of $L$ from left to right and top to bottom. The dark zone correspond to the forbidden region given by $\epsilon<\frac{\mu}{2}$ and the red line corresponds to the horizon. Different colors are used to distinguish different behaviours for the flow. The upper left image is the Penrose diagram for the Kruskal spacetime where the Kerr-Schild patches have been highlighted.}
  \label{fig:FullphasespaceSW}
\end{figure} 
Let us analyze briefly the Schwarzschild dynamical system. The function $h(r)$ takes the form $h(r)=\frac{2M}{r}$. From Theorems \ref{HamiltonT} and \ref{Regularizationtheorem} with the optimal value of $\beta=-\frac{3}{2}$ we derive the geodesic equations
\begin{align}
r'&=r u, \label{SW1} \\
u'&=r \left(L^2-\mu M r\right)-3 L^2 M+\frac{3}{2} u^2,  \label{SW2} \\
\theta'&=L \sqrt{r}.  \label{SW3}
\end{align}

\subsection{The collision manifold}
The submanifold $r=0$ is clearly invariant under the flow. Since $r=0$
corresponds to the spacetime singularity, this submanifold is called
{\it collision manifold}. It can be described
globally by the coordinates $\{u,\theta\}$ so its topology is
$\mathbb{R} \times \mathbb{S}^1$.
The dynamical system (\ref{SW1})-(\ref{SW3}) 
restricted to the collision manifold reads
  \begin{align*}
  u'&=\frac{3}{2} u^2-3 L^2 M, \\
\theta'&=0.
\end{align*}
This system has two lines of critical 
points: one line of stable points at $(\theta,u)=(\theta_0,-\sqrt{2M L^2})$ 
and one line of unstable nodes at $(\theta,u)=(\theta_0,\sqrt{2ML^2})$, where 
$\theta_0 \in \mathbb{S}^1$ is an arbitrary value. For each value of $\theta_0$, there is a 
trajectory extending from $u = - \infty$ and approaching $u=-
\sqrt{2 M L^2}$ as its future limit point, a trajectory from 
$u=- \sqrt{2 M L^2}$ to $u= \sqrt{2M L^2}$ and a trajectory
having $u= \sqrt{2 ML^2}$ as its past limit point and extending
to $u = + \infty$, all of them with $\theta= \theta_0$.

\subsection{The general flow}

We are going to center our attention in the phase portrait for timelike geodesics ($\mu=1$), as this shows more interesting phenomena than 
in the null case. The fixed points are
\begin{equation*}
(r=0, u =\pm \sqrt{2M L^2}), \quad \quad 
(r = r_{\pm}(M,|L|) := \frac{L^2 \pm  |L|\sqrt{L^2-12 M^2}}{2 M}, u=0),
\end{equation*}
the second pair under the additional condition $L^2 \geq 12 M^2$. 
For $L^2 > 12 M^2$ all fixed points are hyperbolic, with $ (r= r_+(M,|L|), u=0)$
being a center (purely imaginary eigenvalues) and $(r=r_{-}(M,|L|), u=0)$
being a saddle. When $L^2 = 12 M^2$, there is a bifurcation in the phase
space, which can be visualized in the transition between the
second and third plots in Fig. \ref{fig:FullphasespaceSW}. We thus recover easily all well-known results for geodesics in Schwarzschild
outside the horizon. The approach here, however, is perfectly regular both
across the horizon at $r=2M$ and even at the singularity $r=0$. Moreover,
it allows us to treat all points in the Kruskal spacetime with a single
dynamical system.

%\begin{figure}
%\centering
%% \hspace*{-0*2.5em}
%\centerline{
%\begin{tabular}{cc}
%\includegraphics[height=0.25\textheight]{collision.png}&
%\hspace{1.5cm}
%\includegraphics[height=0.25\textheight]{cylinder.png}
%\end{tabular}}
%\vspace*{1em}
%\caption{ghvfghgf.}
%  \label{fig:colli}
%\end{figure}

\FloatBarrier
\section{The Reissner-Nordstr\"om dynamical system}

%\begin{figure}
%\centering
%\includegraphics[scale=0.23]{out.png}
%\vspace*{1em}
%\caption{Phase space for massless particles with $ L = 0 $ (left picture) and $ L = 1.5 M $ (right picture). The dark zone correspond to the forbidden region given by $\epsilon<\frac{\mu}{2}$.}
%  \label{fig:bubble}
%\end{figure} 

\begin{figure}
\centering
\includegraphics[scale=0.45]{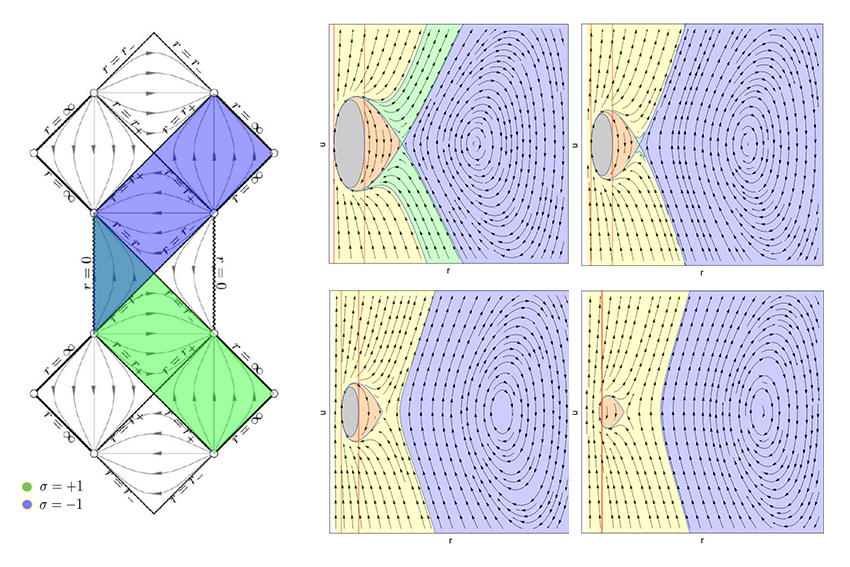}
\vspace*{1em}
\caption{Phase space for timelike particles with fixed value of $L$ increasing value of $Q$ from left to right and top to bottom. The dark zone correspond to the forbidden region and the red line corresponds to the horizons. Different colors are used to distinguish different behaviours for the flow. The left image is the Penrose diagram for the Kruskal spacetime where the Kerr-Schild patches have been highlighted.}
  \label{fig:FullphasespaceRN}
\end{figure} 

Next, we to analyze the Reissner-Nordstr\"om dynamical system. The function $h(r)$ is now $h(r)=\frac{2M r+Q^2}{r^2}$. From Theorems \ref{HamiltonT} and \ref{Regularizationtheorem} with the optimal value of $\beta=-2<\beta_\text{Schwarzschild}$ we derive the geodesic equations
\begin{align}
r'&=r u,  \label{RN1}\\
\theta'&=L \, r,  \label{RN2}\\
u'&=r \left(r \left(L^2-\mu  M r+\mu  Q^2\right)-3 L^2 M\right)+2 \left(L^2 Q^2+u^2\right). \label{RN3}
\end{align}
Unlike the Schwarzschild collision manifold, the Reissner-Nordstr\"om collision manifold is not reachable by timelike geodesics, which reflects the fact that the singularity for charged black holes is repulsive. More details on the approach of causal geodesics to the collision manifold can be found in 
\cite{galindo2014mcgehee}. Concerning the general flow for timelike geodesics, the phase diagram has now three critical points and the excluded region shows
an interesting behaviour: for $0<|Q|<M$ it detaches from the line $r=0$ and moves across the phase diagram, diminishing for larger values of $Q$ and vanishing when $|Q|=M$. As geodesics can now encircle the excluded region, the variation ranges of Lemma \ref{excluded} implies (see 
\cite{galindo2014mcgehee} for details) that any geodesic travelling from $r=r_0> r_+ \to r_+ \to r_{-} \to r_1$ and back to $r_- \to r_+$ must have changed the Kerr-Schild patch along the way (by changing the value of $\sigma$).

It is interesting to note  that inserting 
$Q = 0 $ in system (\ref{RN1})-(\ref{RN3}) the Schwarzschild case is not recovered.  This is because
the value of the parameter $ \beta $ adapted to 
Reissner-Nordstr\"om is different to that of Schwarzschild. Thus,
in the Schwarzschild subcase of Reissner-Nordstr\"om we have overkilled
the singularity and the
fixed points that previously existed at $ r= 0$, $u = u_{\pm}$
have both collapsed to $ u =0$. This collapse
can be detected directly on the Reissner-Nordstr\"om 
phase space  because the fixed point $\{ u=0, \theta=\theta_0\}$
is no longer hyperbolic when $Q=0$.

\section{References}

\bibliography{Bibliography}

\end{document}